\documentclass[aps,prl,twocolumn,groupedaddress,floats,showpacs]{revtex4}

\usepackage{bm}

\usepackage{amssymb,amsmath}
\usepackage{latexsym}
\usepackage{graphicx}
\usepackage{epsfig}

\begin{document}

\preprint{APS/123-QED}

\title{Non--Heisenberg Spin Dynamics of Double-Exchange
Ferromagnets with Coulomb Repulsion}

\author{M. D. Kapetanakis and I. E. Perakis}

\affiliation{Department of Physics, University of Crete, 
and 
Institute of Electronic Structure \& Laser, Foundation
for Research and Technology-Hellas, Heraklion, Crete, Greece} 

\date{\today}

\begin{abstract}
With a variational three--body calculation
we study the role  of the interplay between the onsite Coulomb,
Hund's rule, and  superexchange interactions    
on the  spinwave 
excitation spectrum of itinerant ferromagnets. 
We show that 
correlations between 
a Fermi sea electron--hole pair  and  a 
magnon result in
a very pronounced zone boundary softening 
and strong deviations 
from  the Heisenberg spinwave dispersion.
We  show that this spin dynamics 
depends sensitively on the Coulomb and exchange  interactions and 
discuss its possible relevance 
to recent experiments in the manganites.

\end{abstract}

\pacs{75.30.Ds, 75.10.Lp, 75.47.Lx }
\maketitle

The interaction between itinerant carrier spins and 
localized magnetic moments 
leads to ferromagnetic order 
 in a wide variety  of systems 
\cite{nagaev}.
Examples include 
the manganese oxides (manganites)
R$_{1-x}$A$_{x}$MnO$_{3}$ ( R=La, Pr, Nd, Sm, 
$\cdots$ and  A= Ca, Ba, Sr, Pb, $\cdots$)
 \cite{manganites}
and  the III-Mn-V ferromagnetic 
semiconductors  \cite{III-V}. 
Such systems are of great current 
interest due to their novel potential applications. 
For example, the manganites 
display 
colossal magnetoresistance  \cite{manganites},
while ferromagnetic semiconductors 
raise the possibility 
of multifunctional quantum devices that combine
information processing and  storage
on a single chip with low power consumption \cite{wolf}.
In such  materials, the magnetic and 
transport properties are intimately related and 
can be controlled by varying  
the itinerant carrier concentration
and dimensionality.

In the manganites,
$n$=1-$x$ 
itinerant electrons per Mn atom
partially fill 
 a d--band with e$_{g}$ symmetry. Their concentration, $n$, 
is controlled 
by the hole doping, $x$.  
The d--band kinetic energy $K$
is determined by the hopping energy 
between the neighboring lattice sites, 
$t\sim 0.2-0.5$eV. 
The itinerant electron spins interact strongly 
 with localized 
spin--S magnetic moments
(Hund's rule coupling
$H_{exch}$ with strength $J\sim 2$eV$>t$).
$S$=3/2 comes from the 
three electrons in the tightly bound t$_{2g}$ 
orbitals.
This ferromagnetic interaction competes
with the direct  antiferromagnetic interactions
($H_{AF}$) between neighboring local spins,  
$J_{AF} \sim 0.01$t. 
The largest energy scale in the manganites is given by
the on--site Coulomb  repulsion
 between the itinerant  electrons, $U\sim 3.5-8$eV ($H_U$).
This Coulomb interaction is generally  difficult to treat and 
its effects have received 
less attention. 
Here we focus on the role of $U$ on 
the spin dynamics in the concentration range 
$0.5 \le n \le 0.8$ 
where metallic behavior is observed 
in both  3D and quasi--2D (layered) manganites.

The  ferromagnetic 
order in the manganites can be 
interpreted to first approximation  by 
invoking the double exchange mechanism 
and the  $J \rightarrow \infty$ 
 limit
of the minimal Hamiltonian $K+H_{exch}$ 
 \cite{de,manganites}.
An itinerant carrier is    allowed to hop on a lattice site only if  
its spin is parallel to the local spin on that site. 
The kinetic energy 
is thus reduced when all  spins are  parallel. This 
favors the 
ferromagnetic  state 
$|F\rangle$, which describes 
local spins with $S_z=S$ on all lattice sites 
and a Fermi sea of spin--$\uparrow$  electrons.
The above  spins are often treated as classical, 
justified for 
$S\rightarrow\infty$ \cite{manganites}.
In this limit, 
the system can be 
described by   a
nearest neighbor Heisenberg model with ferromagnetic interaction. 
Quantum effects are often treated 
perturbatively in 1/S \cite{golosov,furukawa}.
To $O(1/S)$, one thus obtains 
noninteracting Random Phase Approximation (RPA) magnons,  
whose dispersion in the strong coupling limit 
coincides with that 
of the nearest neighbor Heisenberg ferromagnet
\cite{furukawa}.
Such a dispersion 
was  observed experimentally
for concentrations  $n>0.7$ 
\cite{sw-exp-1}.

However, strong deviations 
from the short range  Heisenberg model  spinwave dispersion  
were observed 
for  $n\le0.7$
in both 3D 
\cite{soft-exp-3D,endoh,ye} 
and quasi--2D manganites \cite{soft-exp-2D}. 
Most striking is the strong  spinwave softening 
close to the 
zone boundary
 \cite{soft-exp-3D,endoh,ye}. 
This 
indicates a new spin dynamics 
in the metallic ferromagnetic 
phase whose 
 physical origin  
is still unclear  \cite{ye}.  
The proposed mechanisms involve
orbital degrees of freedom,
 magnon--phonon
interactions,  disorder, bandstructure effects, and  the 
 Hubbard repulsion
\cite{ye,endoh,mech,golosov,sun,manganites}.
The zone boundary softening can be 
fitted phenomenologically by 
adding long range interactions
to the Heisenberg Hamiltonian \cite{ye,endoh}.  
Ye {\em et.al.} \cite{ye} found that
the above  softening increases 
with $x$=1-$n$, while 
the dispersion for low momenta only  
changes  weakly.
They argued that 
none 
of the existing theories can explain these
 experimental trends \cite{ye}.

In this paper we study the 
concentration dependence of 
the spinwave dispersion
predicted by the model Hamiltonian 
$H=K+H_{exch}+H_{U}+H_{AF}$ \cite{manganites,golosov,kapet} with a single 
$e_g$ orbital per lattice site.  
We treat exactly the long--range 
magnon--Fermi sea pair three--body  
correlations 
induced by the interplay 
between $H_U$ and $H_{exch}$
with a variational wavefunction.
We show that such correlations lead to strong deviations 
from the RPA 
and Heisenberg spinwave dispersions. 
These deviations, as well as 
the stability of the ferromagnetic order,  
depend sensitively 
on $H_U$.
Our approach  interpolates  between the strong/weak coupling and $n$=0/$n$=1 
limits with the same formalism and 
can therefore address 
the intermediate 
interactions and $n$
relevant to the manganites. 
At the same time, it recovers the 1/S expansion \cite{golosov} 
and 
exact numerical results 
\cite{num,igarashi} 
as special cases. 
We find that  magnon--Fermi sea pair  correlations 
due to $U$ result in 
a pronounced zone boundary spinwave softening
that increases with
$x$ (similar to the 
experiment \cite{ye})  
in a way that depends  on $U$ and$J$. 
Our  variational calculation 
 sets a lower bound on the magnitude of this
softening.

{\em Method---} We use the 
variational wavefunction $|{\bf Q} \rangle =
M^{\dag}_{{\bf Q}} | F \rangle$, where the operator $M^{\dag}_{{\bf
Q}}$ conserves the total momentum ${\bf Q}$ and  lowers the z--component
of the total spin by 1. 
This spin reversal can be achieved either by lowering the localized
spin z--component, via the collective spin operator ${\bf
S}^{-}_{{\bf q}}$ \cite{kapet}, or by coherently promoting an electron from the
spin--$\uparrow$ to the spin--$\downarrow$ band; it 
 may also  be accompanied by the scattering of Fermi sea
pairs. Neglecting  multipair excitations, the 
most general $M_{{\bf Q}}^\dag$ is \cite{kapet}

\begin{figure}[t]
\includegraphics[width=8.7cm]{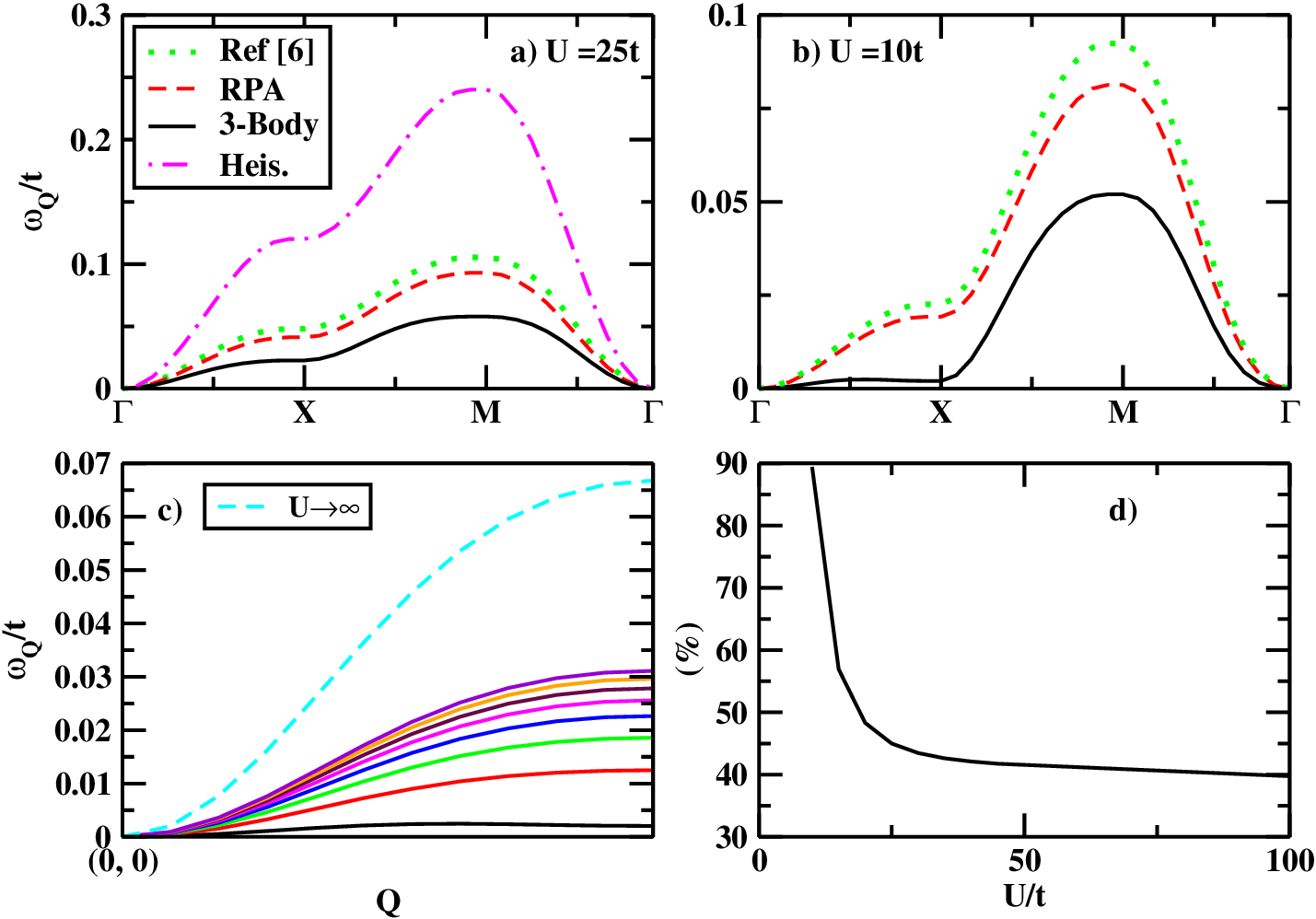}
\caption{(Color online) 
Spinwave dispersion along different directions 
( $n$=0.6, $J$ =7$t$, $J_{AF}$ =0.012$t$).  
(a) U=25t (b) U=10t (c) Direction $\Gamma-X$: 
U=10$\rightarrow$45t in increments of 5t. 
(d) Deviation from the RPA: $1-\omega/\omega^{RPA}$ at $X$--point.
 \label{fig1:n0.6}
}
\end{figure}

\begin{eqnarray} 
\label{magnon}
&& M_{{\bf Q}}^\dag = S_{{\bf Q}}^{-}
      + 
 \sum_{{\bf \nu}} 
  X_{{\bf \nu}}^{{\bf Q}} \
      c_{{\bf Q}+ {\bf  \nu} \downarrow}^{\dag}
      c_{{\bf \nu} \uparrow} 
+       \sum_{{\bf \alpha} {\bf \mu}} 
      c_{{\bf \alpha} \uparrow}^{\dag}
     c_{{\bf \mu} \uparrow} 
\times 
 \nonumber \\
      & &
\left[ 
\Psi^{{\bf Q}}_{{\bf \alpha} {\bf \mu}} \ 
S_{{\bf  Q} + {\bf \mu} - {\bf \alpha}}^{-}
\right. 
+ \left. \frac{1}{2}\sum_{{\bf \nu}}
       \Phi^{{\bf Q}}_{\alpha \mu \nu} 
      c_{{\bf Q}+  {\bf \mu} - {\bf \alpha} + {\bf \nu} \downarrow}^{\dag} 
      c_{{\bf \nu} \uparrow} \right]
  \end{eqnarray}
  where 
 $c^\dag_{{\bf k} \sigma}$ 
creates a spin--$\sigma$, momentum--${\bf k}$ electron.
$\nu , \mu$ ($\alpha$) label states inside (outside) the Fermi sea.
The first two terms 
create  a magnon of momentum ${\bf Q}$. 
The last two terms describe
magnon  scattering, 
${\bf Q}\rightarrow  
{\bf Q}+ {\bf \mu} - {\bf \alpha}$,
accompanied by electron scattering 
across the Fermi surface, 
$\mu\rightarrow\alpha$ (Fermi sea pair shakeup). 
By setting $ \Psi= \Phi=0$ 
we recover the RPA results \cite{kapet}.
However, here the variational parameters  $ X^{{\bf Q}}_{{\bf \nu}}
  ,\, \Psi^{{\bf Q}}_{{\bf \alpha \mu}}$ and  
 $\Phi^{{\bf Q}}_{{\bf \alpha \mu \nu}}$
 are not restricted in any way; 
unlike in Ref. \cite{var},
 we do not 
assume any particular 
form or momentum dependence.
By solving the  full variational equations numerically 
for fairly large $N\times N$ lattices
( $N \sim$ 20-30), 
we put an upper bound on the 
spinwave excitation energies  $\omega_{{\bf Q}}$ 
(with respect 
to  $| F \rangle$)
that  converges with N and thus reflects the thermodynamic limit. 
We can therefore conclude that ({\bf i}) the exact dispersion 
is at least as soft as our results, ({\bf ii}) 
$\omega_{{\bf Q}}<0$  
means that 
$| F \rangle$ is {\em not} the ground state.

The wavefunction Eq.(\ref{magnon}) offers several  advantages.
It gives exact results  
in the two concentration 
 limits $n\rightarrow 0$ (one electron)   and 
$n=1$  (half--filling).
In the special cases  $H_U=H_{AF}=0$ and $H_{exch}=H_{AF}=0$
it agrees very well with 
 exact results 
\cite{kapet,igarashi}. 
Our results  also become exact in the 
atomic limit $t \rightarrow 0$ \cite{kapet,rucken} 
and should therefore treat local correlations 
well.
While  the latter 
dominate in the 
strong coupling limit, long range 
correlations  become important
as  $J/t$ and $U/t$ decrease
 \cite{rucken}.  
The experiment \cite{endoh,ye}  points out the 
importance of  long--range interactions. 
Eq.(\ref{magnon})  treats {\em exactly} all 
correlations between a single Fermi sea pair
and a magnon.
The only restriction
of Eq.(\ref{magnon}) 
 is that it neglects contributions from 
two or more  Fermi sea pairs, which are 
 however suppressed 
for large $S$ \cite{kapet} and in 
1D \cite{rucken,igarashi}.

{\em  Results---} 
Fig.\ref{fig1:n0.6} 
shows the 
calculated  three--body spinwave dispersion
for $U$=25$t$
(Fig.\ref{fig1:n0.6}(a)) 
 and $U$=10$t$ (Fig.\ref{fig1:n0.6}(b)).
It compares this to the RPA  ($\Psi=\Phi=0$) 
and  the results of Ref.\cite{golosov}, which we recover 
by expanding the RPA  to $O(1/S)$  and $O[1/(JS + nU)]$. 
 Fig.\ref{fig1:n0.6}(a) also compares to 
the Heisenberg dispersion 
obtained by taking the limit $J \rightarrow \infty$, $U=0$ of the RPA 
(rather than by fitting). The latter  
deviates strongly from our intermediate coupling results. 
While the RPA  agrees well 
with Ref.\cite{golosov}, the 
Fermi sea pair--magnon correlations
lead to a very strong softening (deviations $\sim$100\% from the RPA).

The on--site 
Coulomb repulsion $U$ increases  the spinwave  energies 
and therefore the stability of the ferromagnetic state $|F\rangle$. 
Fig. \ref{fig1:n0.6}(c) demonstrates this hardening 
along $\Gamma-X$ ((0,0)$\rightarrow$($\pi$,0)) 
as $U$ increases in steps  $\Delta U=5t$. 
While initially 
the energies increase 
strongly with $U$, 
their relative change decreases with 
increasing $U$. Nevertheless, full convergence
 to the $U\rightarrow\infty$ result (dashed curve in 
Fig.\ref{fig1:n0.6}(c)) only occurs for very large $U$. 

 Ref.\cite{sun} treated 
the effects of strong  $U$ by mapping the problem 
to a Hamiltonian with $U=0$ 
\cite{kapet} and  renormalized hopping $t(n)$.
The magnon excitations were then described within the RPA.
Due to the increase in the effective $J/t(n)$, $U$ resulted in 
higher spinwave energies.
Here we show that carrier--magnon correlations beyond the RPA, 
induced by $U$, 
lead to a pronounced zone boundary softening as compared to 
the  RPA. 
This can  be seen in 
 Fig. \ref{fig1:n0.6}(d), which shows the percentage deviation 
from the RPA at the $X$--point 
 as function of $U$
(maximum  is 
100\%).
While the deviations from the RPA  decrease 
with increasing $U$, they remain quite large for the typical $U$.

\begin{figure}[t]
\includegraphics[width=8.7cm]{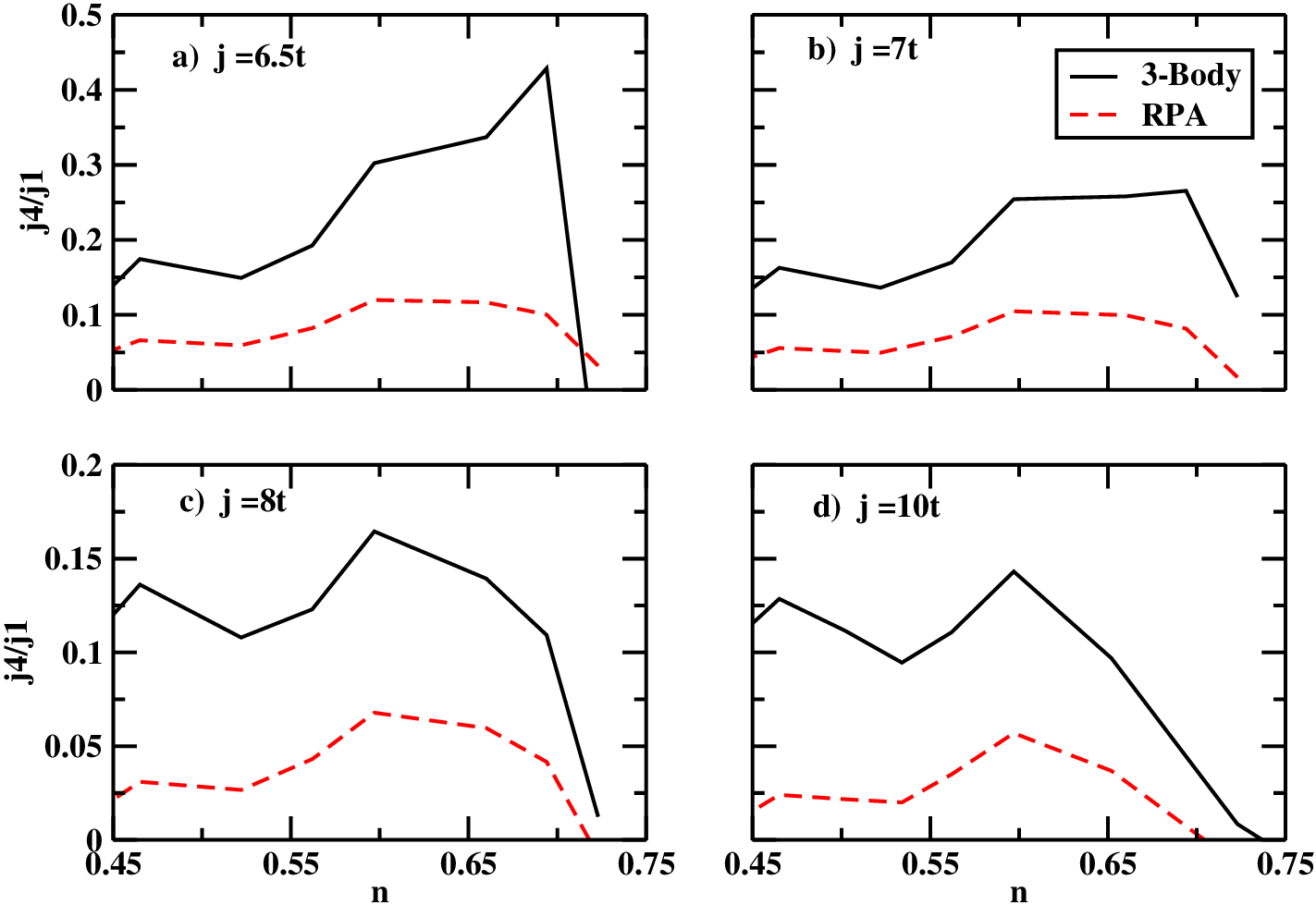}
\caption{
(Color online) 
$J_4(n)/J_1(n)$ for $J_{AF} =0.012t, U =25t$ extracted by fitting 
our results to the 1st+4th nearest neighbor Heisenberg model.
The same behavior is  exhibited by the spinwave 
 softening  compared to the Heisenberg model. 
 \label{fig3:dev}
}
\end{figure}

We now focus on the  dependence of the 
X--point energy on $n$.
Refs.\cite{soft-exp-3D,endoh,ye} 
found that 
the deviation, at this zone boundary, 
of the nearest--neighbor Heisenberg model dispersion
that fits the experiment 
at small  $Q$
increases with 
$x$=1-$n$. 
The experimental dispersion
along all
directions in the Brillouin zone 
was fitted by  
a Heisenberg model with 
{\em both} 4th--nearest--neighbor ($J_4$) and 
next--nearest--neighbor 
($J_1$)
exchange couplings; 2nd-- and 3rd--nearest--neighbor 
interactions were negligible
\cite{ye}.  
The ratio
$J_4/J_1 \propto x$
becomes strong for $n \le 0.7$ \cite{ye}.

Our numerical results can also be fitted very well 
to the $J_1$--$J_4$ Heisenberg model. 
Fig.\ref{fig3:dev}
shows the behavior of $J_4(n)/J_1(n)$ (and thus the spinwave softening) 
for different $J$.
The crucial role of the 
pair--magnon correlations is clear 
by comparing to the RPA. 
The RPA gives   small 
$J_4/J_1$ (in the strong coupling limit it coincides with 
the nearest--neighbor Heisenberg dispersion \cite{kapet}). 
However, the pair--magnon correlations
greatly enhance 
$J_4/J_1$ (and the softening), typically by a factor 3-4 or higher in 
Fig. \ref{fig3:dev}.
$J_4/J_1$ increases rapidly 
with  $x$=1-$n$ 
until it reaches its maximum.
For large $J/t$,
$J_4/J_1$ increases more slowly with $x$.   
This increase is sharp for smaller $J$, 
as the ferromagnetic state becomes less stable (compare 
Figs. \ref{fig3:dev}(a) and \ref{fig3:dev}(d)).
On the other hand, $J_4/J_1$ is small for  $n>$0.7. 

\begin{figure}[t]
\includegraphics[width=8.7cm]{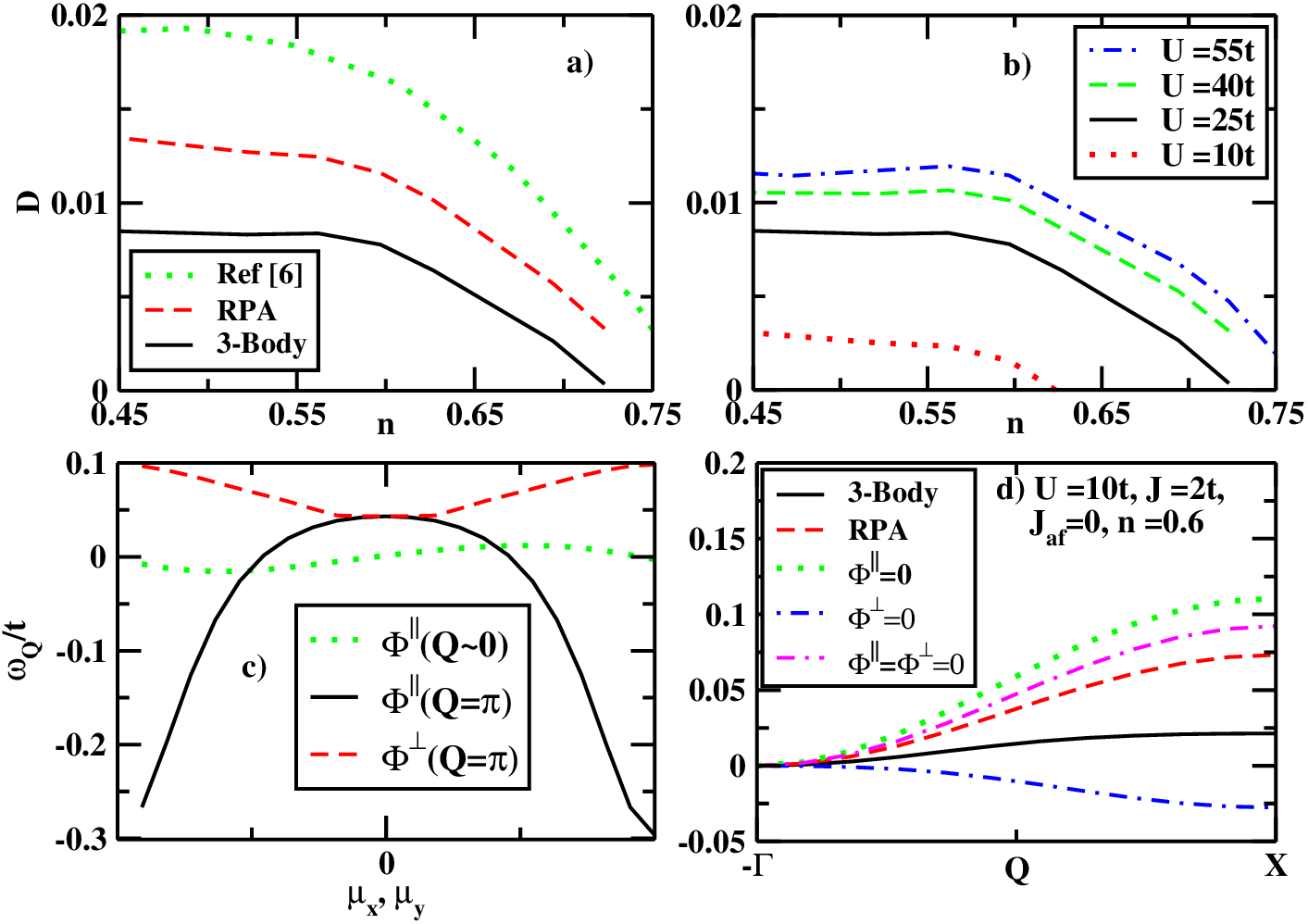}
\caption{
(Color online)  
(a) Comparison of  different approximations
for $D(n)$
($J$=7t, $U$=25$t$),  (b) the role of $U$ on $D(n)$,  
(c) Contribution of  magnon--pair correlations for different 
momenta, 
(d) Origin of magnon softening. $J_{AF}$=0.012$t$.
 \label{fig2:stiffness}
}
\end{figure}

Next we turn to the  spinwave dispersion
for small 
$Q$. 
Its behavior is characterized 
by the stiffness $D(n)$, obtained by fitting the small--$Q$ 
dispersion to the form $D Q^2$. 
Fig.\ref{fig2:stiffness}(a) 
compares our results to Ref.\cite{golosov}, 
Eq.(\ref{magnon}), and the RPA.
The pair--magnon  
correlations decrease $D(n)$ 
by as much as 
$\sim$100\% as compared to Ref.\cite{golosov}
and by as much as  $\sim$50\%
from the RPA.
 Fig.\ref{fig2:stiffness}(a) demonstrates 
a  plateau  as  function of $n$, where $D(n)$ 
remains fairly constant  within a wide range of $n$
relevant to the manganites.
The  pair--magnon  
correlations decrease the dependence of $D$ 
on $n$ for such concentrations
(compare the three curves in Fig.\ref{fig2:stiffness}(a)).  
As shown in Fig.\ref{fig2:stiffness}(b), $U$ increases the stiffness.
Overall,  
 Figs.\ref{fig3:dev} and \ref{fig2:stiffness} 
are
consistent with the main experimental trends \cite{endoh,ye}. 
However, in Ref.\cite{ye} $D(n)$ was found to be fairly constant 
over a wider range of $n$.
 Fig.\ref{fig3:dev} and Fig.\ref{fig2:stiffness} show that the  pair--magnon 
correlations  suppress the dependence of $D$ on $n$ 
while enhancing $J_4$. We speculate that the differences from the experiment 
may be due to the bandstructure effects neglected here.

We now turn to the origin of the zone boundary softening and 
show that it is dominated by strong correlations 
due to $U$.
We set 
$J_{AF}=0$.  
Similar to Ref.\cite{kapet}, 
the spinwave dispersion $\omega_{{\bf Q}}$ 
is determined by the 
amplitude $X^{{\bf Q}}_{\mu}$, Eq.(\ref{magnon}), 
describing the coherent 
 spin$\uparrow$$\rightarrow$spin$\downarrow$ 
electron excitation ($\propto \sum_{\mu} X_{\mu}$) and by
the amplitude $\Psi$,  describing 
magnon--pair scattering.
The dominant new effect here comes from the renormalization of 
$X_{\mu}$ 
by the scattering, due to $U$,  of 
a spin$\uparrow$$\rightarrow$spin$\downarrow$
excitation with a  Fermi sea
pair. 
The corresponding interaction process is described by the amplitude 
$\Phi$ in Eq.(\ref{magnon}) 
and is shown schematically in Fig. 4(a). 
The Fermi sea pair $(\mu, \alpha)$ is created by interacting with 
the spin$\downarrow$ electron via $U$. 
Such scattering  gives a contribution 
 $\propto 
U \sum_{\alpha \nu} \Phi^{{\bf Q}}_{\alpha \nu \mu}$ 
to $X^{{\bf Q}}_{\mu}$.
In Fig 3(c) we plot this correlation contribution,
both for 
${\bf Q}$ close to the $X$--point and for small ${\bf Q}$, 
as function of momentum ${\bf \mu}$ for $n=0.6$ 
where the softening is pronounced. We consider 
 momenta 
${\mu} \| {\bf Q}$ ($\mu_x$, contribution 
$\Phi^{\|}$)  and momenta
 ${\mu} \bot {\bf Q}$
($\mu_y$, contribution $\Phi^{\bot}$).
As can be seen in Fig.3(c), the largest correlation contribution comes for
$\mu \| {\bf Q}$ close to the Fermi surface 
(which for the concentrations of interest is close to the zone boundary)
and for ${\bf Q}$ close to the zone boundary.
In Fig3(d)  we compare the spinwave energy from the 
full calculation 
with the results obtained by neglecting 
$\Phi^{\bot}$ and/or  
$\Phi^{\|}$.  It is clear that 
the  strong softening of the spinwave dispersion 
as compared to the RPA
comes from $\Phi^{\|}$, i.e. from the 
renormalization of $X_{{\bf \mu}}$
by the 
scattering of a spin$\uparrow$$\rightarrow$spin$\downarrow$ 
 excitation with  a Fermi sea pair for momenta   
$\mu$ along $\Gamma$-$X$.

With decreasing $J/t$, 
the magnon energy for intermediate $n$ 
turns negative at the $X$--point while 
the magnon 
stiffness is still positive.
This variational result
allows us to conclude instability of the ferromagnetic state.
 On the other hand, for small $n$, 
the spinwave energy  first turns negative at the $(\pi,\pi)$ 
point (antiferromagnetic correlations).
Finally, for larger $n$, the spinwave energy
turns negative at small momenta  first,  $D<0$. By 
identifying the minimum values of $J$, $J_c(n)$, where  
$\omega_{{\bf Q}} \ge 0$ for all momenta, we 
can definetely  conclude, due to the 
variational nature of our calculation, that 
the ground state is not ferromagnetic 
for $J<J_c$. 
On the other hand, for $J>J_c$, the stability of $|F \rangle$
is not guaranteed.

\begin{figure}[t]
\includegraphics[width=8.0cm]{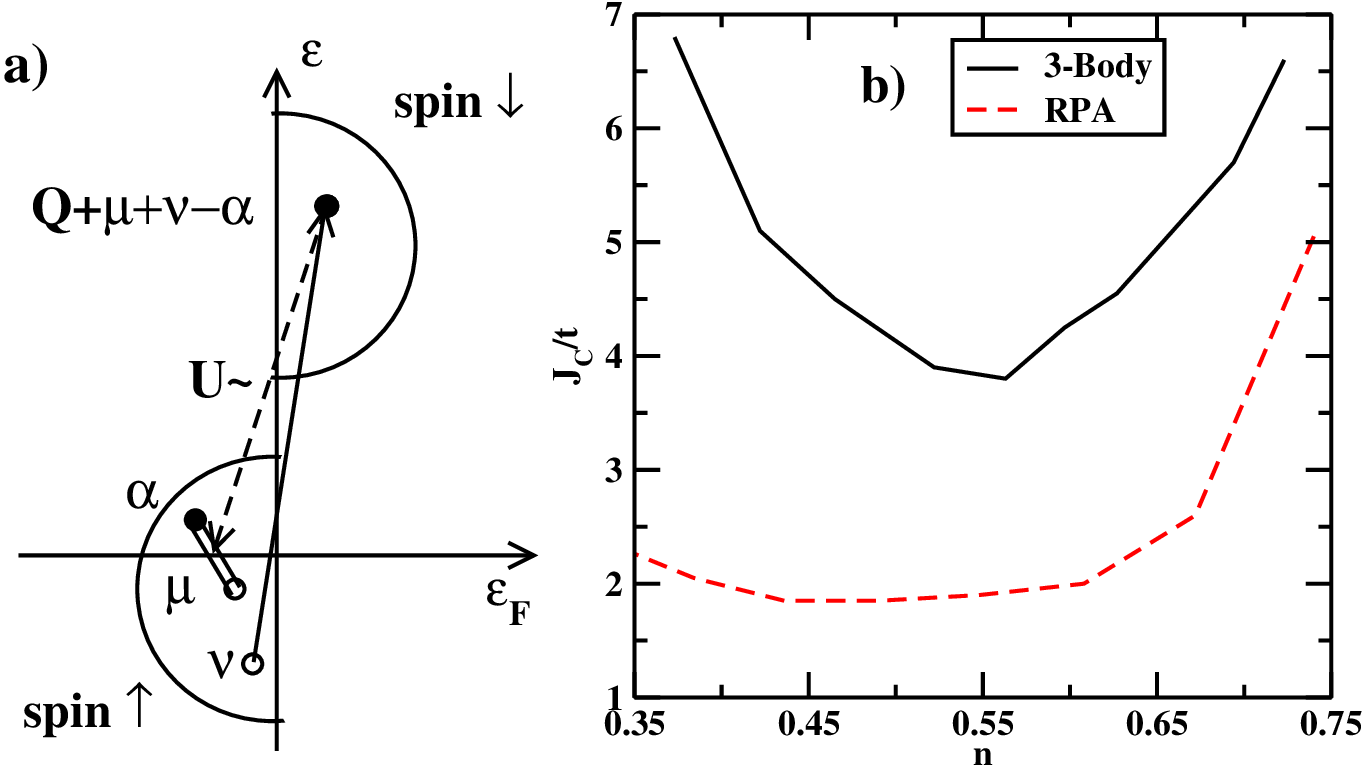}
\caption{
(Color online) (a) Schematic describing the scattering of 
 spin$\uparrow$$\rightarrow$spin$\downarrow$ electronic 
excitation with  Fermi sea
pair  $(\mu,\alpha)$ due to $U$. 
(b) $J_c(n)$ for  
$J_{AF} =0.012t, U =25t$. For 
$J<J_c$, the ferromagnetic state $|F\rangle$ is {\em not} 
the ground state. 
\label{fig4:ph-jvsn}
}
\end{figure}

 $J_c(n)$ is shown in Fig.\ref{fig4:ph-jvsn}(b).
By comparing to the RPA, 
it is clear that the pair--magnon correlations 
lead to a very pronounced upward shift of the ferromagnetic 
phase boundary. 
While for large $n$ the correlation effects 
diminish, and   the RPA becomes exact at $n$=1, 
for $n<$0.7 the deviations from the RPA exceed 100\%. 
As $n$ decreases further, the RPA fails completely and 
we can  conclude 
that it grossly 
overestimates the stability of the ferromagnetism.
Even though
additional effects (e.g. phase separation
\cite{golosov,manganites}
and charge ordering \cite{mishra}) 
will further increase $J_c(n)$ for some $n$,
our variational calculation allows us to 
conclude 
that Fermi sea pair--magnon correlations are strong 
in the manganites and should be 
treated beyond  the mean field theory of Refs. \cite{golosov,mishra}.

We conclude that
 non--perturbative 
long range electron--hole pair--magnon correlations play
a very important role in 
the spin dynamics of the manganites. 
Most important is the strong softening of the 
spinwave dispersion
and the decrease in the stability of the 
ferromagnetic state.
These correlation effects depend sensitively on 
the onsite Coulomb repulsion  and on its interplay 
with the magnetic exchange and superexchange 
interactions.
We propose that the scattering 
of magnons by  charge excitations 
plays an important role in interpreting recent experiments 
\cite{ye}.
Our work can be extended to other itinerant 
 ferromagnetic systems (e.g. 
III(Mn)V  semiconductors) that are far 
from the strong coupling limit.  
The correlations discussed here should also 
play an important role in 
the ultrafast magnetization 
 dynamics measured by pump--probe optical 
 spectroscopy \cite{ultra}. 

M.D.K. acknowledges the support of the
Greek ministry of education and the EU
through the program IRAKLITOS-Research Fellowships. 
I.E.P. acknowledges the support of the  EU program HYSWITCH.

\end{document}